\documentstyle[preprint,prb,aps]{revtex}

\begin{document}
\draft
\title{Electron-electron interactions and two-dimensional\,-\,
two-dimensional tunneling}
\author{T. Jungwirth}
\address{Institute of Physics ASCR,
Cukrovarnick\'{a} 10, 162~00 Praha 6, Czech Republic\\}
\author{A.H. MacDonald}
\address{Department of Physics, Indiana University,
Bloomington, IN 47405}
\date{Received \today}
\maketitle
\begin{abstract}

We derive and evaluate expressions for the dc tunneling conductance
between interacting two-dimensional electron systems at non-zero
temperature. The possibility of using the dependence of the tunneling
conductance on voltage and temperature to determine the
temperature-dependent electron-electron scattering rate at the Fermi
energy is discussed. The finite electronic lifetime produced by
electron-electron interactions is calculated as a function of
temperature for quasiparticles near the Fermi circle. Vertex corrections to
the random phase approximation substantially increase the electronic
scattering rate.  Our results are in an excellent quantitative
agreement with experiment.

\end{abstract}

\section{Introduction}

The development of high mobility double-quantum-well
structures and techniques to independently contact the
two wells even when separated by only tens of nanometers,
have together opened a new
fruitful area of research in low-dimensional physics.  For example,
inter-layer electron-electron interactions have been investigated through the
frictional drag voltage occurring when charge in one layer is moved
relative to charge in the nearby layer \cite{gram,pric}.
Separately contacted double-well structures have turned out to be
extremely useful not only in studies of properties unique to bilayer
systems but also for examining some properties of a single-layer
2D electron gas. In the
case of a gated heterostructure with remotely spaced
quantum wells, where the
inter-layer interactions could be safely neglected, Eisenstein {\it et al.}
\cite{eiscom} were able to relate the electric field leaking between layers
in response to the voltage change on a remote gate
to the compressibility of the electron layer closer to the gate.
2D-2D tunneling\cite{smol,eismag} provides another example where
double-well systems
can be used to probe the electronic properties of individual
electron layers in new ways.
An important feature of ideal 2D-2D tunneling
is the conservation of electron momentum; for 3D-3D tunneling the component
of momentum in the direction of the tunneling barrier is not
conserved.  Momentum conservation is not perfect because of
disorder in the tunneling barrier and in the electron layers, and
also because of inelastic electron-phonon and electron-electron
scattering.  An electron conserving its energy
{\it and} momentum can tunnel only when the subband edges of the two layers are
precisely aligned, resulting in large peak-to valley ratios in the observed
resonant tunneling peaks.   Experimental studies in which the Fermi surfaces
of the 2D electron systems were mapped by measuring the
tunneling conductance in magnetic fields applied parallel
to the 2D planes\cite{eismag} have provided, arguably, the most striking
demonstration that electron momentum is conserved to a remarkable degree in
GaAs/AlGaAs double-well structures.  The heights of resonant peaks in
2D-2D tunneling conductances are limited by scattering processes
which do not conserve the momenta of individual electrons.  At
zero temperature, peak heights and widths may be used to measure
the elastic scattering rate due to disorder.\cite{zhen}
It has been suggested\cite{mur,shayegantun} that the temperature
dependence of 2D-2D tunneling conductance peak heights and widths
can be used to probe inelastic electron-electron or electron-phonon
scattering processes in the individual 2D layers.
By comparing with the measured mobility of their system,
Murphy {\it et al.} \cite{mur} concluded that electron-phonon scattering
could not account for their observations
and suggested that the experiment provides
information predominantly about electron-electron scattering rates.

In this article we discuss the connection between electron-electron
scattering rates in individual layers and 2D-2D tunneling.\cite{zhengdassarma}
We find in Section II that the energy dependence of the
electron-electron scattering rate makes the relationship
to tunneling conductance more complicated than in the case of
elastic disorder scattering.  The formal expressions derived
in Section II nevertheless allow the tunneling conductance to be calculated
from the energy-dependent quasiparticle lifetime.
The problem of calculating the quasiparticle lifetime due to Coulomb
interactions is a standard one in the many-body theory of the
electron gas; for most purposes the random-phase-approximation (RPA) is
reasonably accurate.  For the two-dimensional electron gas, however,
confusing disagreements exist among various analytic evaluations
of the RPA expressions.\cite{chap,hod,giul,fuk}
To clarify the situation, we present in Section III a detailed
derivation of the approximate analytic
formulas for the temperature and energy dependence of the electron-electron
scattering rate.  We follow a line similar to that described, in detail, in
Ref. \onlinecite{hod} and emphasize the points where present work departs
from previous studies.
In Section III, we also discuss calculations of the quasiparticle lifetime
which go beyond the RPA by
including local-field corrections to the effective electron-electron
interaction.  The corrections approximately account for density
and spin-density correlations present in the ground state of
the interacting electron system.
When they are included in lifetime calculations
along with the energy-dependence of the electron-electron scattering rate
, excellent agreement is obtained, as we discuss in Section IV, with the 2D-2D
tunneling experiments of Murphy {\it et al.}.\cite{mur}
In Section V, we  briefly  summarize our results.
An account of a preliminary version of this work has been
presented previously.\cite{ep2ds}


\section{2D-2D tunneling conductance for ${\bf T \ne 0}$}

We consider
a GaAs/AlGaAs heterostructure with two identical quantum wells
and equal layer densities. For typical separations between the 2D-layers
($\sim {\rm 300 \AA} $) inter-layer interaction effects, including
the screening of the intra-layer Coulomb potential by electrons
from the opposite layer, are weak.  All such effects
will be ignored in following
calculations.  The Hamiltonian can be written as the sum of three
terms:

\begin{eqnarray}
\label{ha}
H&=&H_R+H_L+H_T  \nonumber \\
H_T&=&-\sum_{\vec{k},\vec{k}'}\left( t_{\vec{k},
\vec{k}'}c_{\vec{k},R}^{+} c_{\vec{k}',L}+{\rm h.c.} \right)\,.
\end{eqnarray}
$H_R$ and $H_L$ are the Hamiltonian for isolated electrons in
right and left wells including, in general,
contributions from intra-layer interactions and from disorder in
each layer.
The tunneling Hamiltonian, $H_T$, couples the two systems; we assume that
this term can be treated at leading order in perturbation theory.
For tunneling barriers which are invariant under translations
perpendicular to the barrier $t_{\vec{k},\vec{k}'}$ is zero for
$\vec{k}\neq \vec{k}'$ and is independent of $\vec k$;
$t_{\vec{k},\vec{k}'}=t\delta_{\vec{k},\vec{k}'}$.
For noninteracting electrons $t$ determines the
difference in energy between symmetric and antisymmetric
combinations of subband states for the two-layers.

The Kubo formula, which we will use to calculate the tunneling current $I$,
treats the $H_T$ to leading order in perturbation theory.\cite{tunnel}
For 2D-2D tunneling our theory is valid to leading
order in $2 t \tau / \pi \hbar $ where $\tau$ is the lifetime of
an electron in the individual layers.  For noninteracting 2D-electrons
in disorder free double-quantum-well systems
it is never valid to treat $H_T$ as a perturbation.
The condition for the validity of the weak perturbation assumption
is that the mean time for an electron
to move between layers ($\tau_t =   \pi \hbar / 2 t $),
is much longer than the lifetime of electrons due to scattering
within a layer, $\tau$.
An electron hopping from one well
to the other will then scatter, {\it i.e.} change its in-plane momentum,
many times before it jumps back to the first well.

Following a familiar line\cite{tunnel,mah} of reasoning we obtain

\begin{equation}
\label{tc1}
I(V)=\frac{2e}{\hbar} t^2 S \int \frac{d^2
k}{(2\pi)^2}\int_{-E_F}^{\infty} \frac{dE}{2\pi}
\, A\left(E,\vec{k} \right)
A\left(E+eV,\vec{k}\right)
 \left[ n_F\left( E\right) -n_F\left( E+eV\right) \right]
\, ,
\end{equation}
where $S$ is the area of the two-dimensional electron systems,
$eV$ is the difference between chemical potentials in the right and
left quantum wells, $n_F(E)$ is the
Fermi distribution function and $A(E,\vec{k})$ is the spectral
function related to the retarded Green's function by

\begin{equation}
\label{sf}
A(E,\vec{k})=-2{\rm Im} G_{ret}(E,\vec{k})=
\frac{-2{\rm Im} \Sigma_{ret}(E,\vec{k})}{\left[E-
\xi_{k}-{\rm Re}\Sigma_{ret}(E,\vec{k})\right]^2+
\left[{\rm Im}\Sigma_{ret}(E,\vec{k})\right]^2}\; .
\end{equation}
We choose to measure energies from the Fermi energy so that
$\xi_{k}=\hbar^2 k^2/2m -E_F$.
Near the quasiparticle peak the spectral function can be
approximated by a Lorentzian:

\begin{equation}
\label{sfqp}
A(E,\vec{k})=
\frac{ \Gamma(\xi_{k},\vec{k})}{\left(E-
\xi_{k}\right)^2+
\left(\Gamma(\xi_{k},\vec{k})/2\right)^2}\; .
\end{equation}
Here we have neglected the real part of the self-energy which
leads to a physically unimportant rigid shift in the quasiparticle
energies, causes the quasiparticle effective mass to differ
slightly from its free electron value, and slightly reduces the
weight of the quasiparticle peak in the spectral function.  These
effects play a minor role in 2D-2D tunneling experiments\cite{remarkreal}
and we neglect them here in favor of the main effect which comes from the
broadening of the quasiparticle pole.
The width of the Lorentzian peak is related
to the self-energy by $\Gamma(\xi_{k},\vec{k})\equiv -2{\rm Im}
\Sigma_{ret}(\xi_k,\vec{k})$.

In the limit of $\Gamma(\xi_{k},\vec{k})
\rightarrow 0$, corresponding to the noninteracting disorder-free
2D-electron gas, Eq.(\ref{tc1}) can be written as

\begin{equation}
\label{tcf}
I(V)=\frac{2e}{\hbar} t^2 S \frac{g_0}{2}
\int_{-\infty}^{\infty} \frac{dx}{2\pi} \delta(x)\delta(x+eV)
\int_{-E_F}^{\infty} d\xi_k \left[n_F\left( x+\xi_k\right)
-n_F\left( x+\xi_k+eV\right)\right]\; ,
\end{equation}
where $g_0=m/\pi\hbar^2$ is the free-particle density of states
for a 2D electron system and $x\equiv E-\xi_k$.
Because of the $\delta$-functions in Eq.(\ref{tcf}) the integral over
$\xi_k$ gives $eV$ and we obtain for $|eV|\ll E_F$
that the tunneling conductance $G(V)\equiv I(V)/V$ is proportional
to $\delta (V)$. This sharp voltage-dependence
of the tunneling conductance is a direct consequence of electron energy
{\it and} momentum conservation during the tunneling.  An electron with
kinetic energy $\xi_k$ can tunnel only when its potential
energy is conserved, {\it i.e.},
when the energy levels in the quantum wells are
aligned.  For identical wells this condition is
satisfied only at zero voltage.

Scattering processes lead to an uncertainty in the energy of
an electron with a given momentum $\vec{k}$ resulting in a finite width
of the spectral function and broadened peaks in the $G(V)$ curve.
At zero temperature
the broadening is dominated by elastic scattering.  Replacing the
spectral width $\Gamma(\xi_{k},\vec{k})$ in Eq.(\ref{sfqp}) by
$\hbar/\tau_{el}$, where $\tau_{el}$ is the constant elastic scattering
lifetime, Eqs. (\ref{tc1}) and (\ref{sfqp}) give for $\hbar/\tau_{el}\ll E_F$
and $|eV|\ll E_F$:

\begin{equation}
\label{tconc}
G(V)=\frac{2e}{\hbar} t^2 S \frac{g_0}{2}
\frac{2\hbar/\tau_{el}}{(eV)^2+(\hbar/\tau_{el})^2} \; .
\end{equation}
We define $\Gamma_G$ to be the half width of the $G(V)$ curve.
The dependence of the tunneling parameter $t$ on the 
bias voltage can, and will, be neglected.  To see this note
that $ t \sim \exp (- \kappa d_B)$ where $d_B$ is the width
of the barrier between the quantum wells, $\kappa =
(2 m V_B / \hbar^2)^{1/2}$ gives the decay rate of the wavefunction
in the barrier, and $V_B$ is the barrier height.  When a bias
potential is applied the average value of $V_B$ is changed by
$\sim e V$.  Using $ d \kappa / d V_B \sim \kappa/ V_B $ the 
$t$ should change by a factor of $\sim \exp ( - \kappa d_B (eV)/ V_B)$.
In the experiments to which we refer $V_B \sim 300$ meV and the maximum value of 
$|eV|$ is $ \sim 0.1 E_F \sim 0.5 {\rm meV}$. 
We see that the magnitude of the argument of the exponential 
above is much smaller than one.
It follows from Eq. (\ref{tconc}) that
$\Gamma_G = \hbar/\tau_{el.}$ for a system with only elastic scattering.
It has been established experimentally\cite{shayegantun,mur} that
$\Gamma_G$ is temperature dependent, indicating that
some inelastic scattering process is contributing to the
quasiparticle scattering rate.
Samples with different levels of disorder have $G(V)$ peaks whose
half-widths appear to differ by temperature-independent constants.
This property enables elastic and inelastic
contributions to the scattering rate to be separated experimentally;
all the calculations in this article are for a disorder-free system.
Our calculations will help confirm the experimental analysis of Murphy {\it
et al.} who attributed the temperature dependent broadening of the
peak in the $G-V$ characteristic to electron-electron interactions.
If we approximate the spectral width due to electron-electron interactions,
$\Gamma_{e-e}(\xi_k,T)$, by $\Gamma_{e-e}(0,T)$ we arrive again at
Eq.(\ref{tconc}) with $\hbar/\tau_{el}$
replaced by $\Gamma_{e-e}(0,T)$.   We show later that this
neglect of the energy  dependence of $\Gamma_{e-e}(\xi_k,T)$
underestimates $\Gamma_G$.


\section {Electron-electron scattering rate}

As explained above,
both the temperature dependence and the energy dependence
of $\Gamma_{e-e}$  need to be calculated in order
to compare theory to measured $G-V$ characteristics.
In this section we present calculations for a pure 2D-electron gas in the
random phase approximation and in the local-field-corrected RPA.
The contribution of electron-electron scattering to the
spectral width of the one-particle Greens function
can be expressed
\cite{das,zhengdassarma} in terms of the scattering rates
of electrons ($\hbar/\tau_e$) and holes ($\hbar/\tau_h$):
\begin{eqnarray}
\label{gamma}
\Gamma_{e-e}(\xi_k,T)&=&\frac{\hbar}{\tau_e(\xi_k,T)}+
\frac{\hbar}{\tau_h(\xi_k,T)}\nonumber \\
&{}&\nonumber \\
\frac{\hbar}{\tau_e(\xi_k,T)}&=&
\sum_{\sigma '}\int\frac{d^2k'}{(2\pi)^2}\int\frac{d^2p}{(2\pi)^2}
W^{\sigma\sigma '}n_F(\xi_{p})[ 1-n_F(\xi_{k'})]
[
1-n_F(\xi_{p'})]
\delta(\xi_{k}+\xi_{p}-\xi_{k'}-\xi_{p'})\nonumber \\
&{}&\nonumber \\
\frac{\hbar}{\tau_h(\xi_k,T)}&=&\frac{n_F(\xi_{k})}{\left[
1-n_F(\xi_{k})\right]}\frac{\hbar}{\tau_e(\xi_k,T)}\; .
\end{eqnarray}
In Eq.(\ref{gamma}), \ $\vec{k},\vec{p}$ are the initial electron momenta and
$\vec{k}',\vec{p}\ '$  are the final electron momenta.
Because of the conservation of the total momentum in electron-electron
scattering processes, $\vec{p'}=\vec{k}+\vec{p}-\vec{k'}$.
$W^{\sigma\sigma '}$ is the scattering function which we now discuss.


\subsection{Scattering function in the random phase approximation}

In the RPA, the scattering function $W^{\sigma\sigma '}$ is
spin-independent
($W^{\uparrow\uparrow}_{RPA}=W^{\uparrow\downarrow}_{RPA}\equiv
W_{RPA}$) and depends only on the momentum transfer
$q\equiv |\vec{k'}-\vec{k}| = |\vec {p'} - \vec{p}|$
and energy transfer $\hbar\omega\equiv\xi_{k'}-\xi_{k}=\xi_{p}-\xi_{p'}$:

\begin{equation}
\label{cpe}
W_{RPA}=2\pi\left|\frac{v(q)}{\varepsilon_{\rm RPA}
(q,\hbar\omega)}\right|^2\; .
\end{equation}
In Eq.(\ref{cpe}) $v(q)=g_0^{-1}\,q_{TF}/q$  is the
unscreened Coulomb interaction
and $\varepsilon_{\rm RPA}(q,\hbar\omega)=1-v(q)\chi_0(q,\hbar\omega)$ is
the RPA dielectric function\cite{stern,hod,giul} for the 2D-electron
gas. (Here $q_{TF}=g_0e^2/2\epsilon$ is Thomas-Fermi screening wavevector
and $\chi_0(q,\hbar\omega)$ is the susceptibility
of a noninteracting electron gas \cite{stern,mald,hod,giul}.)

At low temperatures the energy transferred in electron-electron
scattering ($\sim k_B T$) is small  compared to the Fermi
energy and the magnitude of the momentum transfer  is restricted
to the interval $(0,2k_F)$. In this limit the
real part of the susceptibility ${\rm Re} \chi_0(q,\hbar\omega)\approx-g_0$,
the imaginary part ${\rm Im} \chi_0(q,\hbar\omega)\approx 0$, and we can write

\begin{equation}
\label{cpa}
W_{RPA}\approx\frac{2\pi}{g_0^2}\frac{(q_{TF}/q)^2}{(1+q_{TF}/q)^2}\; .
\end{equation}
Previous analytic evaluations of the RPA quasiparticle lifetimes
have employed the approximation $W_{RPA}\approx 2\pi/g_0^2$,
which would be reasonable if $q_{TF}\gg 2k_F$ or
if the slowly varying function $W_{RPA}(q)$ is multiplied, in the
integrand in Eq.(\ref{gamma}), by a function sharply peaked
near $q=0$.  Neither of these assumptions is valid however since:
(i) The condition $q_{TF}\gg 2k_F$ corresponds to the dimensionless parameter
$r_s=(q_{TF}/2k_F)/\sqrt{2}$,
conventionally used to render the density of an electron gas,
being much larger than 1.
At such low electron densities it is known that the RPA
fails.  In fact, in the next section it is shown that even for $r_s\sim
1$ vertex corrections to the RPA substantially increase the electron
electron scattering rate.  Moreover, in GaAs, for electron
concentrations typical of tunneling experiments the Thomas-Fermi screening
wavevector is comparable to the Fermi wavevector; (ii)  As we now
discuss in detail the integrand in Eq.(\ref{gamma}) has
sharp peaks near both forward ($q=0$) and backward ($q=2k_F$) scattering
momentum transfers.

In polar coordinates
$d^2k'=k'\,dk'\,d\theta_{k'}$ and $d^2p=p\,dp\,d\theta_{p}$,
where the  angles $\theta_{k'}$ and $\theta_{p}$
are measured with respect to momentum $\vec{k}$.
The integral over $\theta_{p}$ may be performed using
the energy-conservation $\delta$-function.  Since $\xi_{p'}$
depends on $\theta_{p}$ we may write
\begin{equation}
\label{delta}
\delta (\xi_{k}+\xi_{p}-\xi_{k'}-\xi_{p'})=\sum_{i}\left|\frac{
\partial\xi_{p'}}{\partial\theta_{p}}\right|^{-1}_{\theta_{p}=\theta_{p,i}}
\delta(\theta_{p}-\theta_{p,i})\; ,
\end{equation}
where $\theta_{p,i}$, is an angle at which both the energy conservation

\begin{equation}
\label{ec}
p'^2=k^2+p^2-k'^2
\end{equation}
and momentum conservation
\begin{equation}
\label{mc}
p'^2=(\vec{k}+\vec{p}-\vec{k'})^2
=k^2+p^2+k'^2+2kp\cos \theta_{p}-2kk'\cos \theta_{k'}-2pk'\cos
(\theta_{k'}-\theta_{p})\;
\end{equation}
conditions are satisfied.  From Eq.(\ref{mc}) it follows that

\begin{equation}
\label{pd}
\frac{\partial p'^2}{\partial\theta_{p}}=-2kp\sin\theta_{p}+
2pk'\sin (\theta_{p}-\theta_{k'})=-2pk'\sin\theta_{k'}\frac{\cos (
\theta_{p}+z)}{\cos z}\; ,
\end{equation}
where

\begin{equation}
\label{z}
\tan z=-\frac{k-k'\cos\theta_{k'}}{k'\sin\theta_{k'}}
\end{equation}
The combination of Eqs.(\ref{ec}), (\ref{mc}) and (\ref{z}) gives
\begin{equation}
\label{k'}
k'^2=\frac{pk'\sin\theta_{k'}}{\cos z}\sin (\theta_{p}+z)+kk'\cos\theta_{k'}
\end{equation}
and together with Eq.(\ref{pd}) we finally obtain

\begin{eqnarray}
\label{pdf}
\left|\frac{
\partial\xi_{p'}}{\partial\theta_{p}}\right|^{-1}_{\theta_{p}=\theta_{p,i}}&=&
\frac{1}{2}\left[A+(E_F+\xi_{k})(E_F+\xi_{k'})\sin^2\theta
_{k'}\right]^{-1/2}\nonumber \\
A&=&(\xi_{p}-\xi_{k'})
\left[\xi_{k}+\xi_{k'}+2E_F-(\xi_{k}+E_F)^{1/2}(\xi_{k'}+E_F)^{1/2}
\cos\theta_{k'}\right]
\end{eqnarray}

In the limit of small $T$ and $\xi_{k}$ the Fermi functions restrict energies
(measured from the Fermi energy)
of particles involved in the scattering process to a small region near zero
energy. As seen from Eq.(\ref{pdf}),
$A$ is then small and the integrand in (\ref{gamma}) has equivalent
sharp peaks in the available phase space for scattering
near $\theta_{k'}=0$ and $\theta_{k'}=\pi$.
The main contributions to the electron-electron scattering rate come from
processes with small wavevector transfer
(forward scattering) {\it and} wavevector transfer
$\sim 2k_F$ (backward scattering).  This suggests that
$W_{RPA}(q)$ may be approximated by the average of its value at $q=0$
and its value at $q = 2 k_F$:

\begin{equation}
\label{waa}
W_{RPA}\approx\frac{2\pi}{g_0^2}w^{f,b}_{RPA}\; ,
\end{equation}
where

\begin{equation}
\label{w0}
w^{f,b}_{RPA}=\frac{1+\left(1+\frac{1}{r_s\sqrt{2}}\right)^{-2}}{2}\; .
\end{equation}


\subsection{Scattering function in the local-field-corrected RPA}

In estimating quasiparticle scattering rates the RPA does
not account for electronic correlations in the interacting electron gas.
Technically, the RPA for the electronic self-energy neglects
vertex corrections to the dynamically screened exchange energy.
On physical grounds correlations are expected to suppress
scattering between like-spin electrons since these electrons
are required to avoid each other by the Pauli exclusion principle and
to enhance scattering between opposite-spin electrons.
In general the four-point scattering amplitude for electrons
depends on the energies and momenta of all electrons and not
just on the momentum ($q$) and energy ($\omega$) transferred
in the scattering event.
Nevertheless, a number of workers\cite{otherlf,mac} have suggested similar
approximations in which the four-point scattering wavefunction
is replaced by an effective electron-electron interaction dependent
only on $q$ and $\omega$.  In these approximations correlations
are accounted for by modifying the bare electron-electron interaction
to take account of the correlation clouds carried around by each
quasiparticle.

Here we estimate corrections to the RPA by adopting the effective
electron-electron interaction suggested by MacDonald and Geldart.\cite{mac}
Their effective interaction has density-density
($t_{nn}$) and spin-spin ($t_{mm}$) components:

\begin{eqnarray}
\label{Wsda}
W^{\uparrow\uparrow}_{LFRPA}&=&2\pi(t_{nn}+t_{mm})^2\nonumber\\
W^{\uparrow\downarrow}_{LFRPA}&=&2\pi(t_{nn}-t_{mm})^2\; .
\end{eqnarray}
Here the density-density interaction,

\begin{equation}
\label{tnn}
t_{nn}=\frac{v(q)+F_{nn}(q)}{1-\chi_0(q,\hbar\omega)\left(
v(q)+F_{nn}(q)\right)}\; ,
\end{equation}
is a screened Coulomb interaction and the spin-spin interaction,

\begin{equation}
\label{tmm}
t_{mm}=\frac{F_{mm}(q)}{1-\chi_0(q,\hbar\omega)F_{mm}(q)}\; ,
\end{equation}
can be thought of as exchange interaction which favors parallel spin
alignment.
The local field factors $F_{nn}(q)$ and $F_{mm}(q)$
which appear in Eq.(\ref{tnn}) and Eq.(\ref{tmm}) are related to
the static density ($\chi_{nn}(q)$) and spin ($\chi_{mm}(q)$)
response functions of the electron gas:
\begin{eqnarray}
\label{Fsa}
F_{nn}(q)&=& \chi_0(q)^{-1} - \chi_{nn}(q)^{-1}  \nonumber\\
F_{mm}(q)&=& \chi_0(q)^{-1} - \chi_{mm}(q)^{-1}.
\end{eqnarray}
\, From Eq.(\ref{Fsa}) we see explicitly that the local fields vanish
in the RPA.
Quantum Monte-Carlo calculations of the response functions indicate
that the wavevector dependence of the local fields is weak\cite{cep}
for $q \leq 2 k_F $.  Here we approximate the local fields by their
$q \to 0$ limits which are\cite{yong} accurately known.
Using the same arguments as in the previous section,
the four point scattering function $W_{LFRPA}^{\sigma\sigma '}$ can then be
approximated by the average
between its forward and backward scattering limits, {\it i.e},

\begin{equation}
\label{Wsdaa}
\frac{W_{LFRPA}^{\uparrow\uparrow}+W_{LFRPA}^{\uparrow\downarrow}}{2}=
2 \pi (t_{nn}^2+t_{mm}^2) \approx\frac{2\pi}{g_0^2}\,w^{f,b}_{LFRPA}\; ,
\end{equation}
where

\begin{equation}
\label{wsda}
w^{f,b}_{LFRPA}=\frac{1+\left(1+\frac{1}{r_s\sqrt{2}
+g_0 F_{nn}}\right)^{-2}}{2}+
\left(\frac{g_0 F_{mm}}{1+g_0 F_{mm}}\right)^2\ .
\end{equation}

In Figure \ref{f1} we compare scattering functions in
the random phase approximation and in the local-field-corrected RPA
as a function of density.
The RPA result is recovered in the high-density limit where the
local fields go to zero. For $r_s=1$, corresponding to the electron density
of the sample in the experiment\cite{mur} of
Murphy {\it et al.}, the local-field-corrections increase the scattering
rate by approximately 30\%.  As also shown in Figure \ref{f1},
there is a $\approx 50\%$ difference at $r_s =1$ between the forward scattering
approximation for the scattering function
and the approximation obtained by averaging forward and backward limits.
(In the forward scattering limit $w^f_{RPA}=1$ and
$w^f_{LFRPA}=1+[g_0F_{mm}/(1+g_0F_{mm})]^2$.)


\subsection{Approximate analytic results for low temperatures and energies}

Analytic expressions for the RPA electron lifetime $\hbar / \tau_{e}$
at low energies and/or temperatures
have been derived previously in four independent
studies\cite{chap,hod,giul,fuk} with slightly different results obtained
in each case.  None of these expressions agree with the
analytic results we present below, whose accuracy has been
verified by comparing with independent numerical calculations.
While some of the discrepancies appear to be due to inadvertent
algebraic errors, some are associated with significant aspects of
the physics of electron-electron scattering in two-dimensional
electron systems which we draw attention to below.  Because of
the existing   confusion we give a detailed description of our calculation.
We have found the detailed analysis presented
in Ref. \onlinecite{hod} to be very helpful and have followed this
calculation closely.

We have shown above that the phase space for electron-electron
scattering at low-temperatures is dominated by contributions
with equal weight near the forward and backward scattering limits.
This property invalidates the approximation, made in
previous analytic evaluations, in which the RPA screened
interaction is approximated by its forward scattering limit.
For our analytic calculation we have adopted an approximation
which is in the same spirit by replacing the scattering function
by Eqs.(\ref{waa}),(\ref{w0}) or Eqs.(\ref{Wsdaa}),(\ref{wsda}).
Making this replacement we can take the interaction strength
outside all integrals.  We integrate first
over the angle $\theta_{k'}$ between the incoming and scattered momenta
in Eq. (\ref{gamma}).
Using equations similar to (\ref{delta}-\ref{pdf}) we find that

\begin{equation}
\label{pdf2}
\left|\frac{
\partial\xi_{p'}}{\partial\theta_{k'}}\right|^{-1}_
{\theta_{k'}=\theta_{k',i}}=
\frac{1}{2}\left[(\xi_{k'}-\xi_{p})
(\xi_{k}-\xi_{k'})+(E_F+\xi_{k})(E_F+\xi_{p})\sin^2\theta
_{k'}\right)^{-1/2}\; .
\end{equation}
As illustrated in Figure \ref{f2} there are two different angles,
$\theta_{k',1}$ and $\theta_{k',2}$,
with identical integration weights (\ref{pdf2})
which satisfy the energy and momentum conservation conditions.
The corresponding momenta are related by a mirror
symmetry with respect to the vector $\vec{k}+\vec{p}$.
(The associated factor of two in the scattering rate appears to have
been missed in some previous work.)
At low $T$ we can take

\begin{equation}
\label{app}
(E_F+\xi_{k})(E_F+\xi_{p})\approx E_F^2 \; .
\end{equation}
Defining $\tilde{\xi_k}\equiv\xi_{k}/k_BT$, $\tilde{\xi_{k'}}
\equiv\xi_{k'}/k_BT$,
and $\tilde{\xi_p}\equiv\xi_{p}/k_BT$ leads to

\begin{eqnarray}
\label{gamma2}
\frac{\hbar}{\tau_e(\tilde{\xi_k},T)}&\approx& 4\frac{m^2}{(2\pi\hbar)^4}
\frac{(k_BT)^2}{E_F}\frac{2\pi}{g_0^2}w^{f,b}
\int_{-\infty}^{\infty}d\tilde{\xi_{k'}}\int_{-\infty}^{\infty}
d\tilde{\xi_p}\, n_F(\tilde{\xi_p})\nonumber\\
&\times&\left[1-n_F(\tilde{\xi_{k'}})\right]\left[1-
n_F(\tilde{\xi_k}+\tilde{\xi_p}-\tilde{\xi_{k'}})\right] I(B)\; ,
\end{eqnarray}
where $w^{f,b}$ stands for either $w^{f,b}_{RPA}$ or $w^{f,b}_{LFRPA}$
and
\begin{equation}
\label{B}
B=\left(\frac{k_BT}{E_F}\right)^2(\tilde{\xi_{k'}}-
\tilde{\xi_p})(\tilde{\xi_k}-\tilde{\xi_{k'}})
\end{equation}
The angular integral $I(B)$ reads

\begin{eqnarray}
\label{IB}
I(B)&=&2\int_{\theta_0}^{\pi-\theta_0}\frac{d\theta_p}{2\left(
B+\sin^2\theta_p\right)^{1/2}}\nonumber \\
&=&2\int_{u_0}^1\frac{du}{\left(1-u^2\right)^{1/2}
\left(B+u^2\right)^{1/2}}\; .
\end{eqnarray}
Here $u_0=0$ for $ B \ge 0$ and $u_0 = \sqrt{|B|}$ for $B < 0$.
To find the leading low-temperature behavior
of the electron-electron scattering rate we replace the expression
$\left(B+u^2\right)^{1/2}$ by first two terms of its Taylor expansion and
approximate the elliptic integral Eq.(\ref{IB}) by
\begin{equation}
\label{IBa}
I(B)\approx\ln(8)-\ln|B|\; .
\end{equation}
Inserting Eq.(\ref{IBa}) into Eq.(\ref{gamma2}) and using the identity

\begin{equation}
\label{ff}
\int_{-\infty}^{\infty}d\tilde{\xi_{k'}}
\int_{-\infty}^{\infty}d\tilde{\xi_p}
n_F(\tilde{\xi_p})\left[1-n_F(\tilde{\xi_{k'}})\right]
\left[1-n_F(\tilde{\xi_k}+\tilde{\xi_p}-\tilde{\xi_{k'}})\right]=
\frac{1}{2}(\pi^2+\tilde{\xi_k}^2)\left[1-f(\tilde{\xi_k})\right]
\end{equation}
we arrive at

\begin{eqnarray}
\label{gamte}
\frac{\hbar/\tau_e(\tilde{\xi_k},T)}{E_F}&\approx&
\frac{w^{f,b}}
{\pi}\left(\frac{k_BT}{E_F}\right)^2
\left[1-n_F(\tilde{\xi_k})\right]\nonumber \\
&\times&\left[\frac{1}{2}(\pi^2+\tilde{\xi_k}^2)\left(\frac{\ln 8}
{2}+\ln\left(\frac{E_F}{k_BT}\right)\right)-F(\tilde{\xi_k})\right]\; ,
\end{eqnarray}
where

\begin{eqnarray}
\label{Ft}
F(\tilde{\xi_k})&=&\frac{1}{2}\left[1-n_F(\tilde{\xi_k})\right]^{-1}
\int_{-\infty}^{\infty}d\tilde{\xi_{k'}}
\int_{-\infty}^{\infty}d\tilde{\xi_p}
\ln\left|(\tilde{\xi_{k'}}-\tilde{\xi_p})
(\tilde{\xi_k}-\tilde{\xi_{k'}})\right|\nonumber\\
&\times&n_F(\tilde{\xi_p})\left[1-n_F(\tilde{\xi_{k'}})\right]
\left[1-n_F(\tilde{\xi_k}+\tilde{\xi_p}-\tilde{\xi_{k'}})\right]\; .
\end{eqnarray}

In the limit of $\tilde{\xi_k}\rightarrow 0$, 
{\it i.e.}, for an electron on the Fermi
surface,
$F(0)=0.41388$ and we obtain the leading contribution to the electron-electron
scattering rate at low temperatures

\begin{equation}
\label{gamt}
\frac{\hbar/\tau_e(0,T)}{E_F}\approx
w^{f,b}\frac{\pi}{4}\left(\frac{k_BT}{E_F}\right)^2
\left[\ln\left(\frac{E_F}{k_BT}\right)+\frac{\ln 8}
{2}-.083\right]\; .
\end{equation}
As follows from Eq.(\ref{gamma})
the scattering rates of electrons and holes are
equivalent at $\xi_k=0$, {\it i.e.},

\begin{equation}
\label{gt}
\Gamma_{e-e}(0,T)=2\frac{\hbar}{\tau_e(0,T)}\; .
\end{equation}

At zero temperature the calculation is simplified by the fact
that the Fermi distribution functions are reduced to step
functions.  We find that for $\xi_k>0$

\begin{equation}
\label{gamee}
\frac{\Gamma_{e-e}(\xi_k,0)}{E_F}=\frac{\hbar/\tau_e(\xi_k,0)}{E_F}\approx
w^{f,b}\frac{1}{2\pi}\left(\frac{\xi_k}{E_F}\right)^2
\left[\ln\left(\frac{E_F}{\xi_k}\right)+\frac{\ln 8}
{2}+.5\right]
\end{equation}
and for $\xi_k<0$

\begin{equation}
\label{gameh}
\frac{\Gamma_{e-e}(\xi_k,0)}{E_F}=\frac{\hbar/\tau_h(\xi_k,0)}{E_F}\approx
w^{f,b}\frac{1}{2\pi}\left(\frac{\xi_k}{E_F}\right)^2
\left[\ln\left(-\frac{E_F}{\xi_k}\right)+\frac{\ln 8}
{2}+.5\right]\;.
\end{equation}


\section{Numerical results}
All the calculations we discuss in this section have been performed
for $r_s=1$.
In Figure \ref{f3} the  low-temperature and low-energy
approximate analytic results (Eqs.(\ref{gamt}),(\ref{gt}) and
(\ref{gamee}),(\ref{gameh})) are compared with numerical results obtained
within the constant  interaction approximation
(\ref{Wsdaa}),(\ref{wsda}). For
temperatures or excitation energies up to 10\% of the Fermi temperature or
Fermi energy, resp., the terms with higher powers of $T$ or $\xi_k$ can be
safely neglected.  The discrepancies here are due only to the
low-temperature
and low-energy approximations.  We see that including the $T^2$ and
$\xi_k^2$ corrections to the leading $-T^2 \ln (T)$ and
$-\xi_k^2 \ln |\xi_k|$ terms greatly extends the range of validity
of these expressions. Note that the spectral width due to electron-electron
interactions is {\it not} precisely an even function of energy.
In the constant interaction approximation at zero temperature
the difference between scattering rates of an electron with
energy $\xi_k>0$ and of a hole with
energy $-\xi_k$ is due to the term given by Eq.(\ref{pdf2}). Since
$(E_F+|\xi_k|)>(E_F-|\xi_k|)$ the integrand in Eq.(\ref{gamma}) is smaller for
electrons than for holes with the same absolute value of energy.
Making the approximation
of Eq.(\ref{app}) which only affects terms of higher order in $\xi_k$ than
$\xi_k^2\ln|\xi_k|$ and $\xi_k^2$
gives the leading low-energy behavior which is even in $\xi_k$.

The validity of the constant scattering amplitude approximation
(Eq.(\ref{Wsdaa}) and Eq.(\ref{wsda})) was examined by performing
three different numerical calculations of $\Gamma_{e-e}(0,T)$ and
$\Gamma_{e-e}(\xi_k,0)$. In Figure \ref{f4} we show results obtained
using (i) the constant scattering function
$W^{f,b}_{LFRPA}=2\pi/g_0^2\,w^{f,b}_{LFRPA}$ (These results are
identical to the curves
labeled "Numerical" in Figure \ref{f3}.), (ii) the $q$-dependent approximation
$W^a(q)$ to the scattering function which results
when the wavevector and frequency-dependence of the susceptibility
$\chi_0(q,\omega)$ is neglected ($\chi_0(q,\omega)\approx -g_0)$,
 and (iii) the full $q$ and $\omega$ dependent scattering function
($W_{LFRPA}(q,\omega)$)
in the local-field-corrected RPA where both the real and imaginary parts of
$\chi_0(q,\omega)$ are taken into account.  We see that averaging
between forward and backward scattering limits provides
an excellent approximation for the $q$-dependence of
$W^a(q)$.  The discrepancy between $\Gamma_{e-e}$ calculated with
the constant ($W^{f,b}_{LFRPA}$) and full scattering function
($W_{LFRPA}(q,\omega)$) is primarily due to
neglecting the imaginary part of $\chi_0$ which  is non-zero
at finite frequencies.

At $T \ne 0$ quasielectrons and quasiholes are both present
both above and below the Fermi energy.  The energy-dependent
spectral width $\Gamma_{e-e}(\xi_k)$
is then a sum of electron and hole scattering rates.
Numerical results showing electron and hole scattering rate
contributions are shown
for $T=0.3\,T_F$, in Figure \ref{f5}.
Spectral widths calculated at several different temperatures are plotted
in Figure \ref{f6}. Dotted curves show the results of calculations where the
temperature dependence of the susceptibility $\chi_0$ is neglected.
This approximation greatly simplifies non-zero temperature numerical
calculations, however, we see here that
it can be safely used only for temperatures
$T<0.1\,T_F$.

Inserting $\Gamma_{e-e}(\xi_k)$ into
Eq.(\ref{sfqp}) we can evaluate Eq.(\ref{tc1}) for the tunneling
current and calculate the tunneling conductance as a function
of applied voltage at a fixed temperature.
At low temperatures and energies the spectral width $\Gamma_{e-e}(\xi_k,T)$
is, roughly, the sum of $\Gamma_{e-e}(0,T)$ and $\Gamma_{e-e}(\xi_k,0)$ (see
Figure \ref{f6}). Looking at the approximate analytical formulas
(\ref{gamt})-(\ref{gameh}) we see that if $T/T_F\ll 1$ and $|\xi_k|/E_F\ll 1$
then also $\Gamma_{e-e}(\xi_k,T)/E_F\ll 1$.
For $|eV|/E_F\ll 1$ we can therefore replace
the difference between Fermi functions in Eq.(\ref{tc1})
by $\partial n_F(E)/\partial E\,\times eV$ resulting in
the following formula for the tunneling conductance

\begin{eqnarray}
\label{gva}
G(V)&=&\frac{2e}{\hbar} t^2 S \frac{g_0}{2}
\int_{-\infty}^{\infty} \frac{dE}{2\pi} \int_{-E_F}^{\infty} d\xi_k
\frac{ \Gamma_{e-e}(\xi_{k})}{\left(E-
\xi_{k}\right)^2+
\left(\Gamma_{e-e}(\xi_{k})/2\right)^2} \nonumber \\
&{}& \nonumber \\
&\times&
\frac{ \Gamma_{e-e}(\xi_{k}-eV)}{\left(E+eV-
\xi_{k}\right)^2+
\left(\Gamma_{e-e}(\xi_{k}-eV)/2\right)^2}
\, \frac{\partial n_F(E)}{\partial E} \;.
\end{eqnarray}

Using Eq.(\ref{gva}) we determined numerically the
effect of the energy dependence of the spectral width due to
electron-electron interactions on $\Gamma_G(T)$.
As shown in Figure \ref{f7},
approximating $\Gamma_{e-e}(\xi_k,T)$ by $\Gamma_{e-e}(0,T)$
underestimates  $\Gamma_G(T)$ by a factor $\sim$ 1.1-1.16 in the range of
temperatures used in the tunneling experiment\cite{mur}.
In Figure \ref{f8}, the calculated $\Gamma_G(T)$ is compared to the
measured half-width of the tunneling conductance peak, showing excellent
agreement between the theoretical result obtained in the
local-field-corrected RPA and experimenal data.  Note that the agreement between
theory and experiment has been achieved without introducing any
adjustable parameters.  As for the
scattering rate calculation,
local-field-corrections increase the RPA $\Gamma_G(T)$. At
$T/T_F=0.15$ the correction factor is $\sim$ 1.3 and 
improves agreement between theory and experiment.  While it is 
certainly possible that the excellent agreement with experiment is
partly fortuitous, our results leave little
doubt that the temperature dependence observed in the 2D-2D
tunneling experiments is due to the finite quasiparticle lifetimes
which result from electron-electron scattering.


\section{\bf Summary}

In 2D-2D tunneling experiments, resonant tunneling features
have no width in the absence of quasiparticle scattering
processes.  This property of 2D phase space allows
2D-2D tunneling experiments to measure quasiparticle scattering rates.
In this article we have examined the role of electron-electron scattering
in recent 2D-2D tunneling experiments.  For
elastic scattering the width of the resonant tunneling
peak is proportional to the scattering rate for electrons
at the Fermi energy.  For
electron-electron scattering we find that the relationship
is complicated by the non-negligible
dependence of the scattering rate on quasiparticle
energy relative to the Fermi energy.  We have shown that,
although the energy dependence of the scattering rate
results in a non-Lorentzian lineshape for the resonance,
it has only a modest influence on the temperature dependence of the half-width
of resonant tunneling peak.  Analytic expressions have
been derived which approximate RPA and local-field-corrected RPA results
for the scattering rates at low temperatures and energies.
The accuracy of these analytic expressions has been
confirmed by comparison with independent exact numerical
evaluations of the scattering rates in these approximations.
For the electron densities of existing experiments
local-field-corrections increase the RPA scattering rate
by a factor $\sim$ 1.3 and are important in improving agreement
between theory and experiment.

\acknowledgements

We acknowledge helpful interactions with D.M. Ceperley,
J.P. Eisenstein, Y. Katayama, S.Q. Murphy, M. Shayegan, and L. Zheng.
This work was supported by National Science Foundation under Grants
INT-9106888 and DMR-9416906.

\begin{figure}
\caption{Forward ($f$) and the average between
forward and backward ($f,b$)
limits of the scattering function calculated in the random phase approximation
(RPA) and in the local-field-corrected RPA (LFRPA).}
\label{f1}
\end{figure}

\begin{figure}
\caption{Schematic illustration of angles between the incoming
and scattered momenta for two-dimensional electron-electron
scattering.}
\label{f2}
\end{figure}

\begin{figure}
\caption{Spectral width as a function of temperature at zero energy (upper
panel) and as a function of energy at zero temperature (lower panel)
calculated using the constant scattering amplitude approximations
discussed in the text.  These results are for the
local-field-corrected RPA.  Numerical calculations (full lines)
are compared to these approximate analytic results.}
\label{f3}
\end{figure}

\begin{figure}
\caption{Spectral width as a function of temperature at zero energy (upper
panel) and as a function of energy at zero temperature (lower panel)
calculated numerically
in the local-field-corrected RPA.  Exact LFRPA calculations (full lines)
are compared to results obtained with $q$-dependent approximate
scattering function (dashed lines) and with the constant scattering
amplitude (dotted lines).}
\label{f4}
\end{figure}

\begin{figure}
\caption{Spectral width (full line) as a sum of electron and hole scattering
rates (dotted lines) at $T=0.3\,T_F$.}
\label{f5}
\end{figure}

\begin{figure}
\caption{Spectral widths (full lines) calculated at different temperatures.
Dotted lines show results obtained using the zero temperature analytic form
of free electron susceptibility.}
\label{f6}
\end{figure}

\begin{figure}
\caption{Half-width at half-maximum
of the tunneling conductance peak relative to the spectral
width at $\xi_k=0$. The inset shows the tunneling conductance at
$T=0.05\,T_F$ (dotted line) and at $T=0.3\,T_F$ (dashed line) and the
tunneling conductance calculated with $\Gamma_{e-e}(\xi_k,T)$ replaced by
$\Gamma_{e-e}(0,T)$. The energy dependence results in a line-shape
for the 2D-2D tunneling resonance which is not precisely Lorentzian.}
\label{f7}
\end{figure}

\begin{figure}
\caption{Half-width at half-maximum
for 2D-2D resonant tunneling peaks: theoretical results including 
local-field corrections to the RPA (full
line), theoretical results in the RPA (dashed line),  
Murphy {\it et al.} experiment (dotted line).}
\label{f8}
\end{figure}

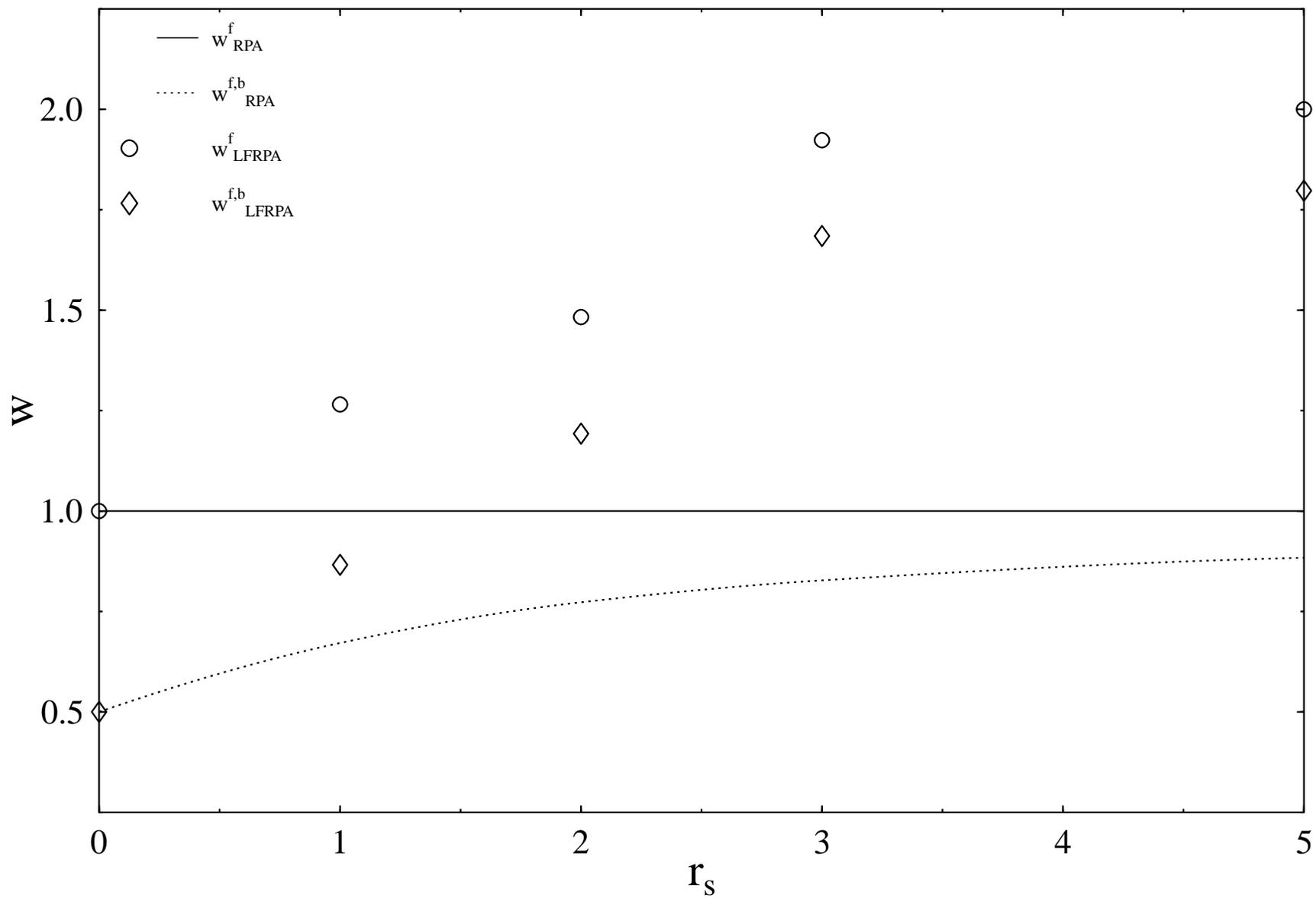
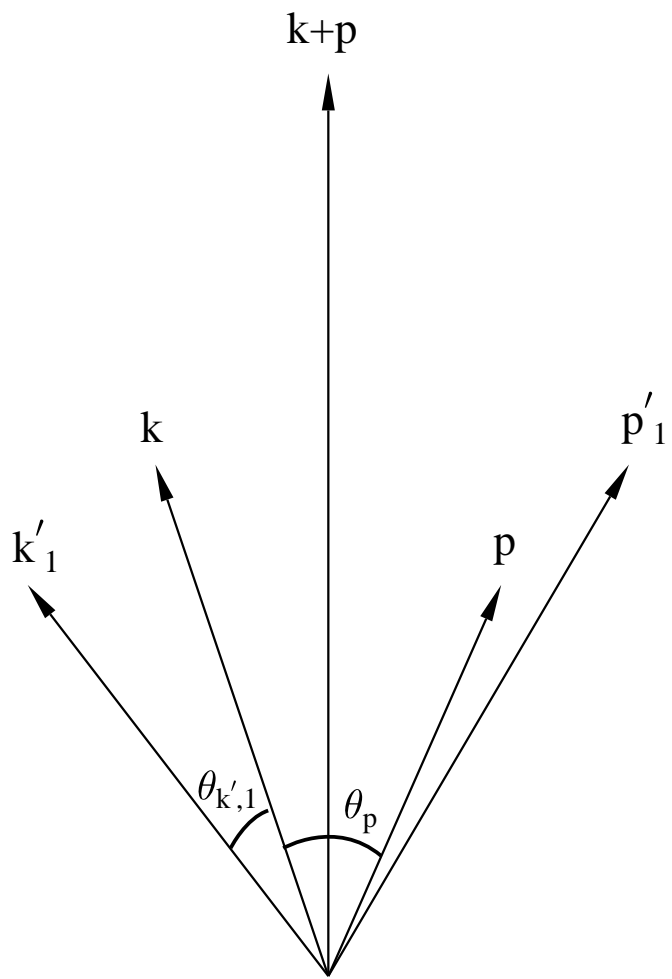
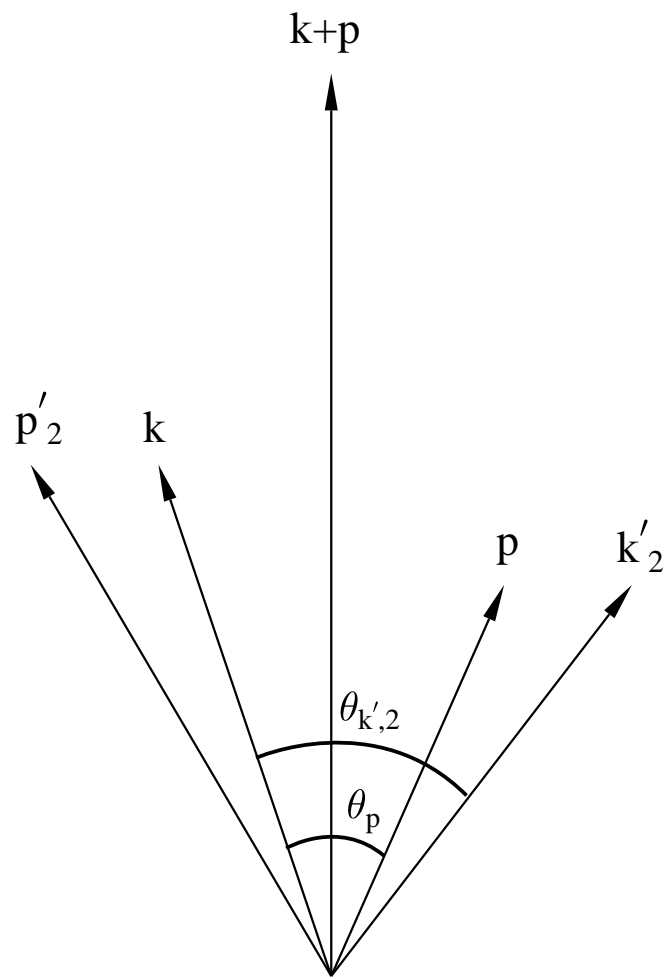
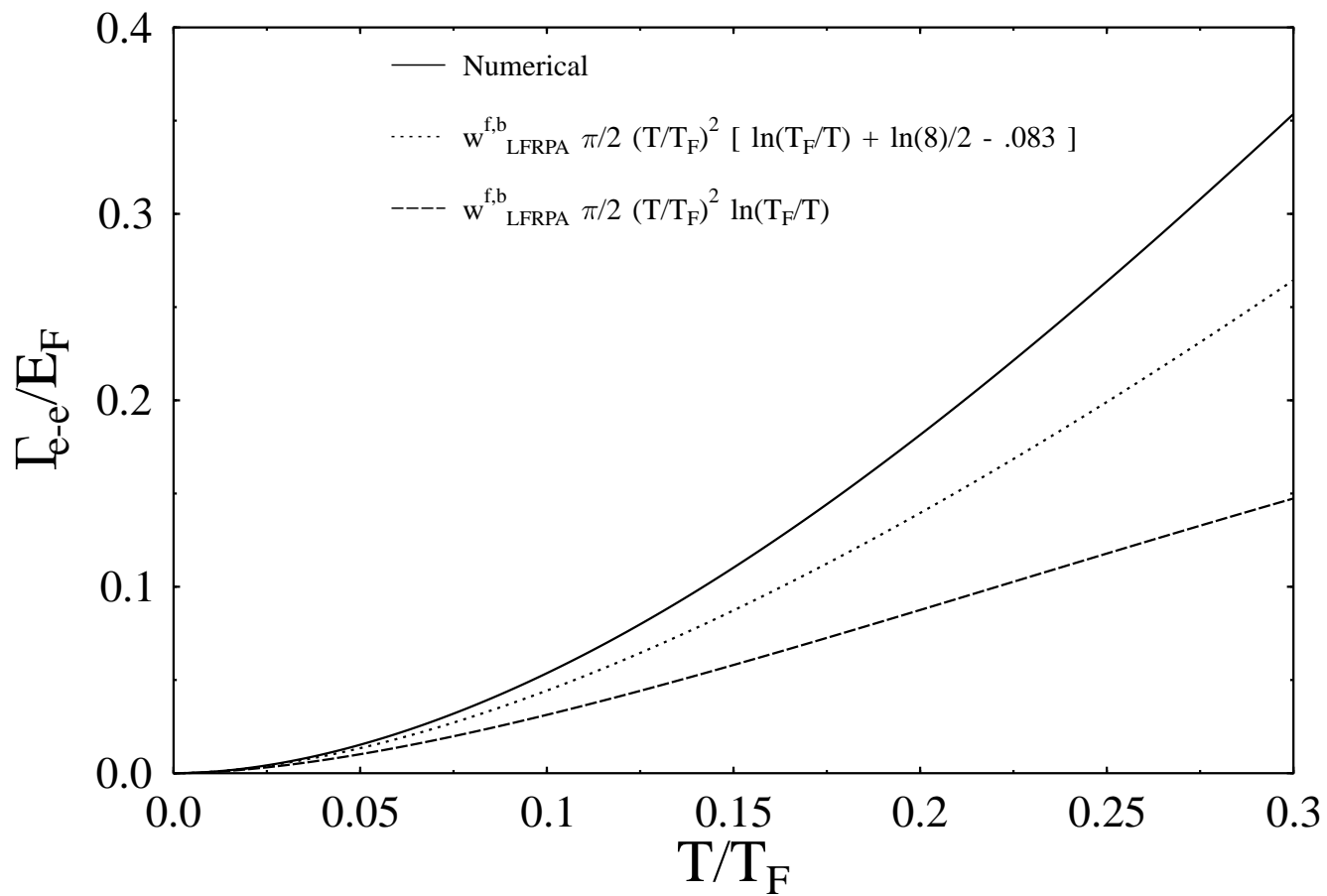
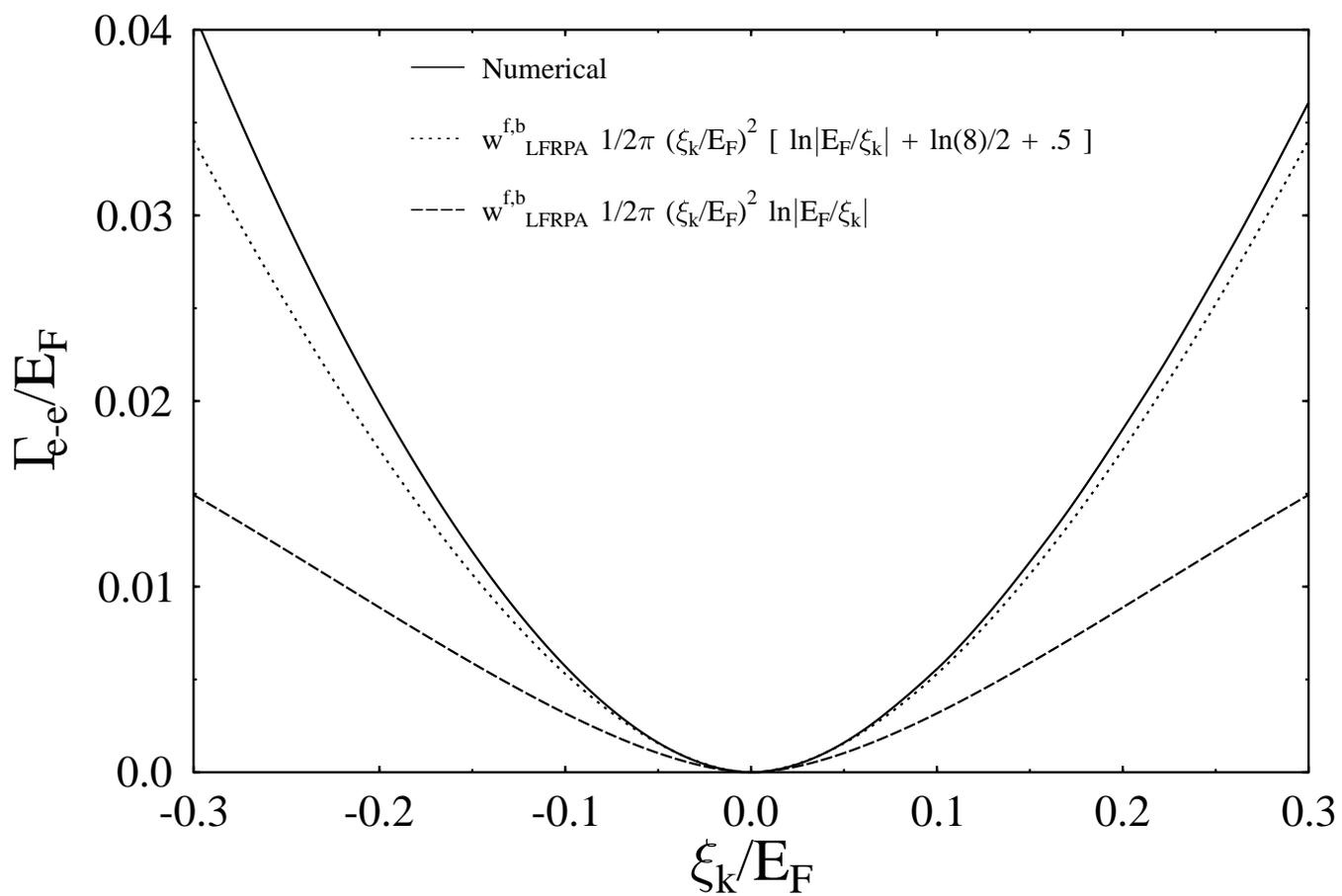
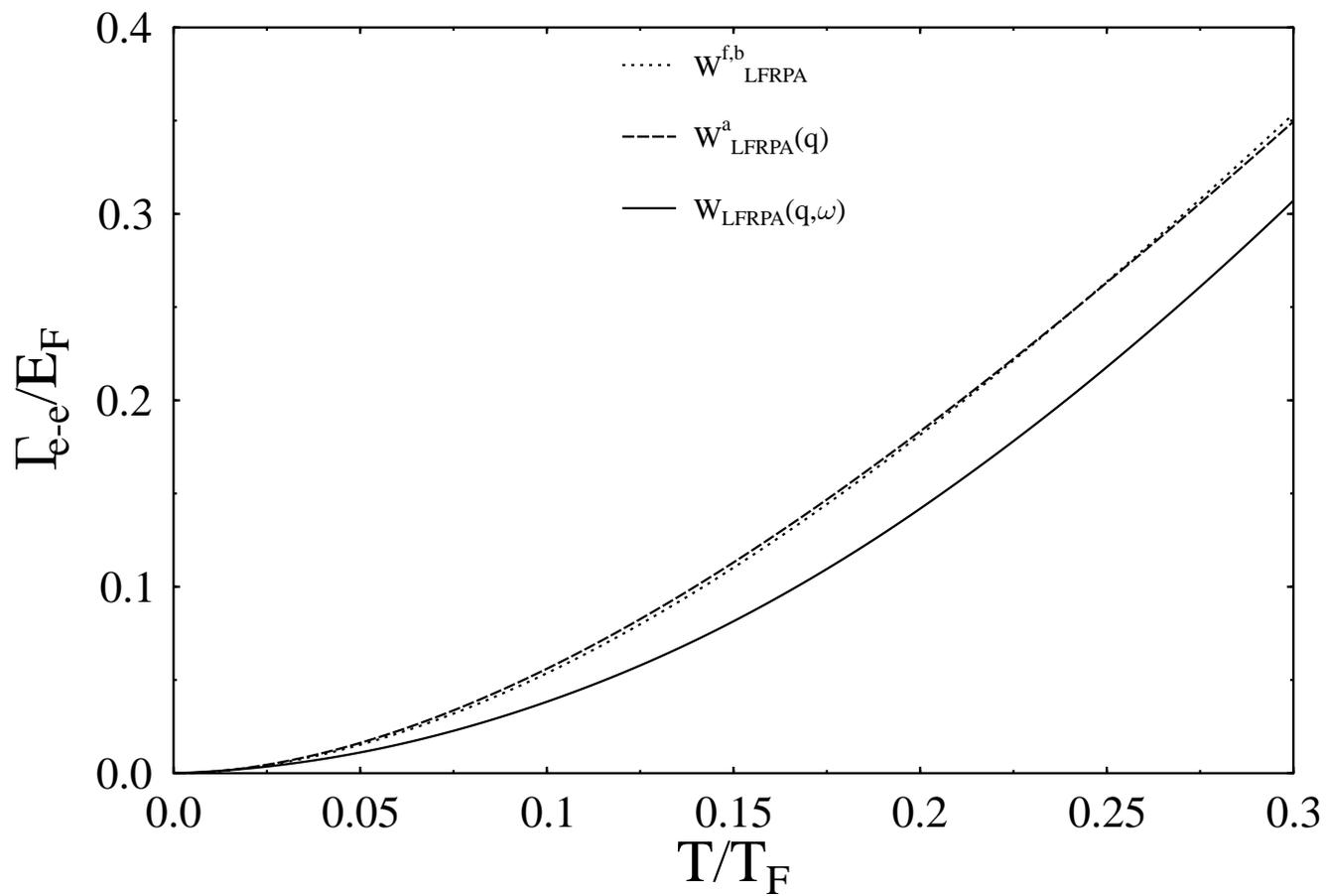
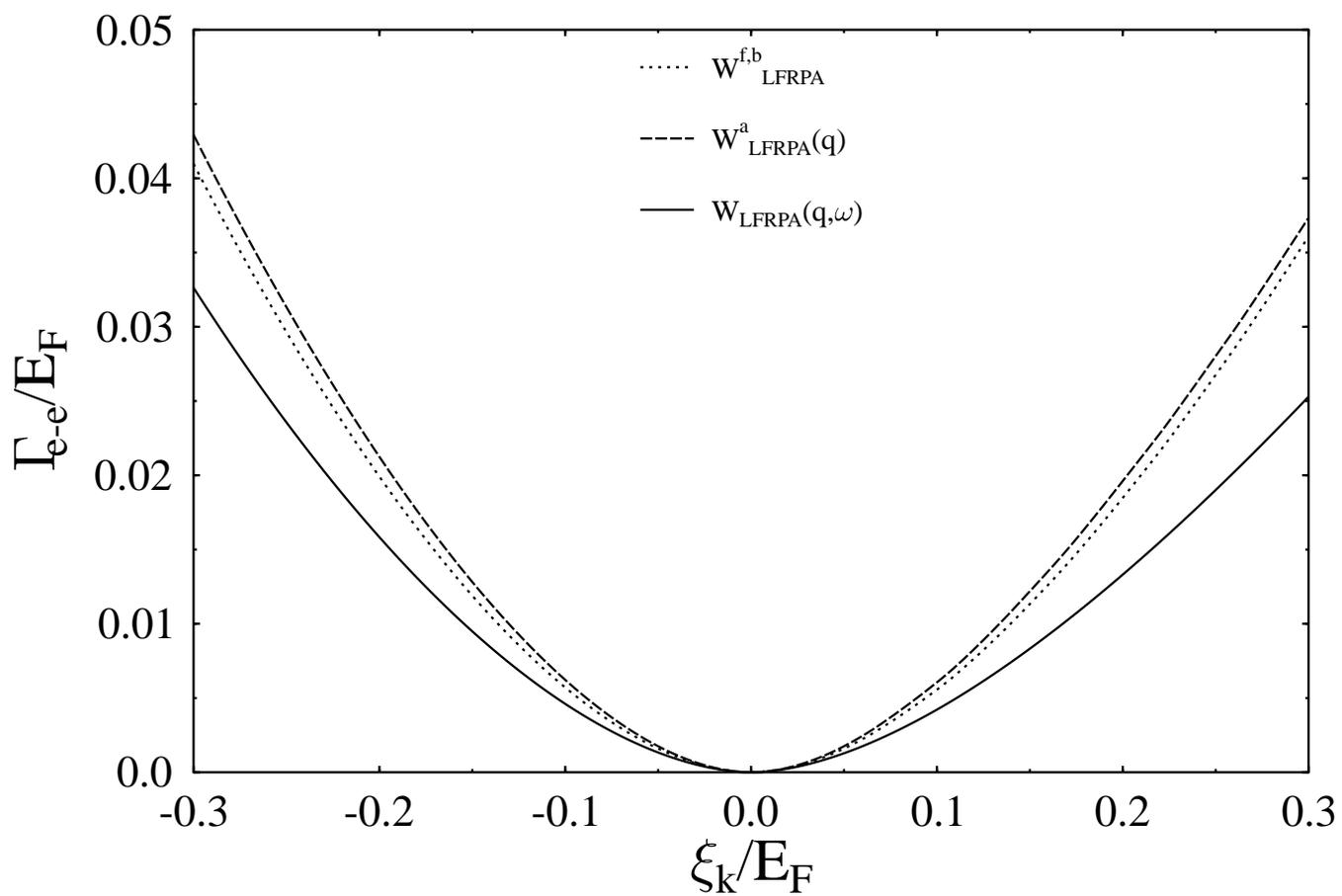
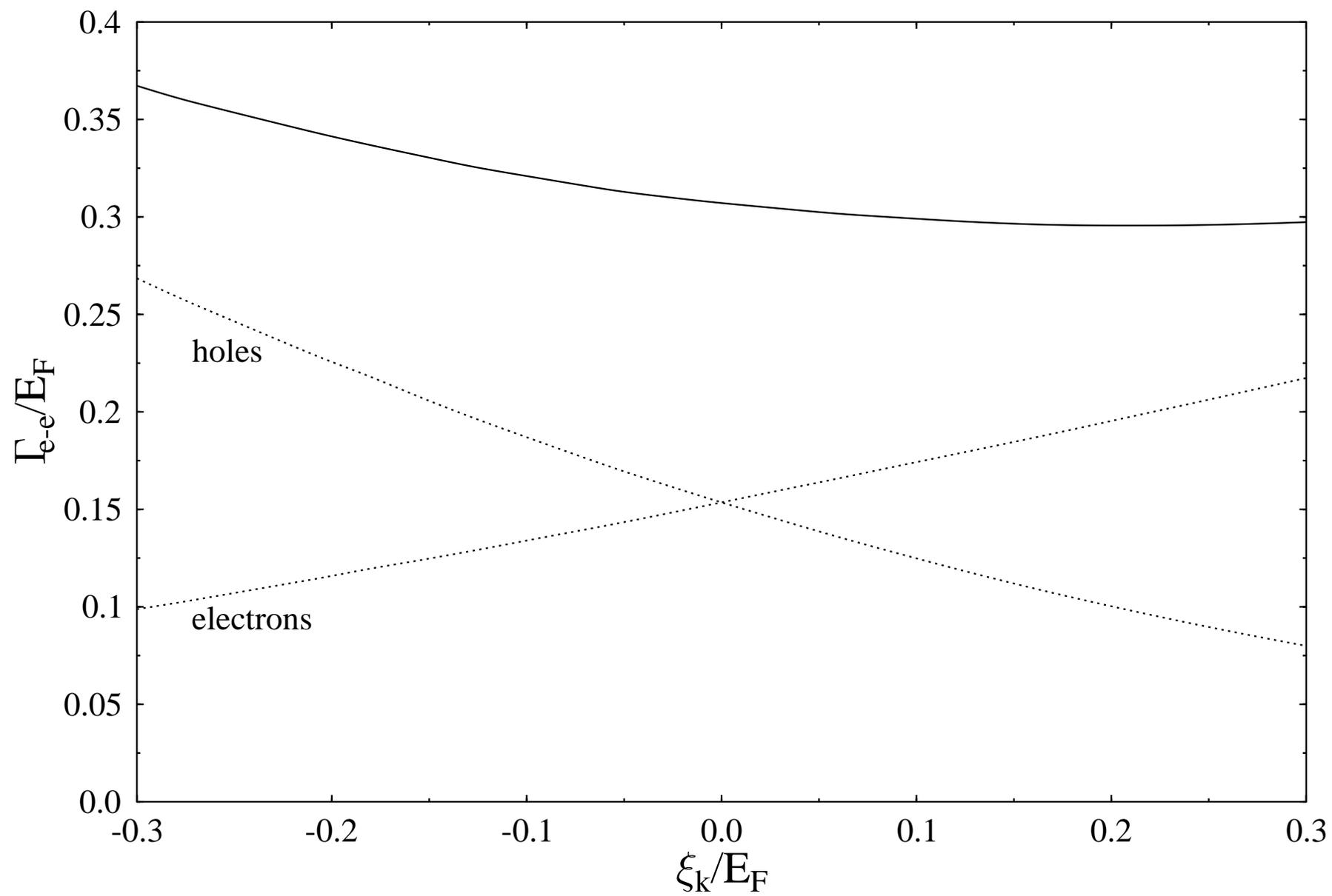
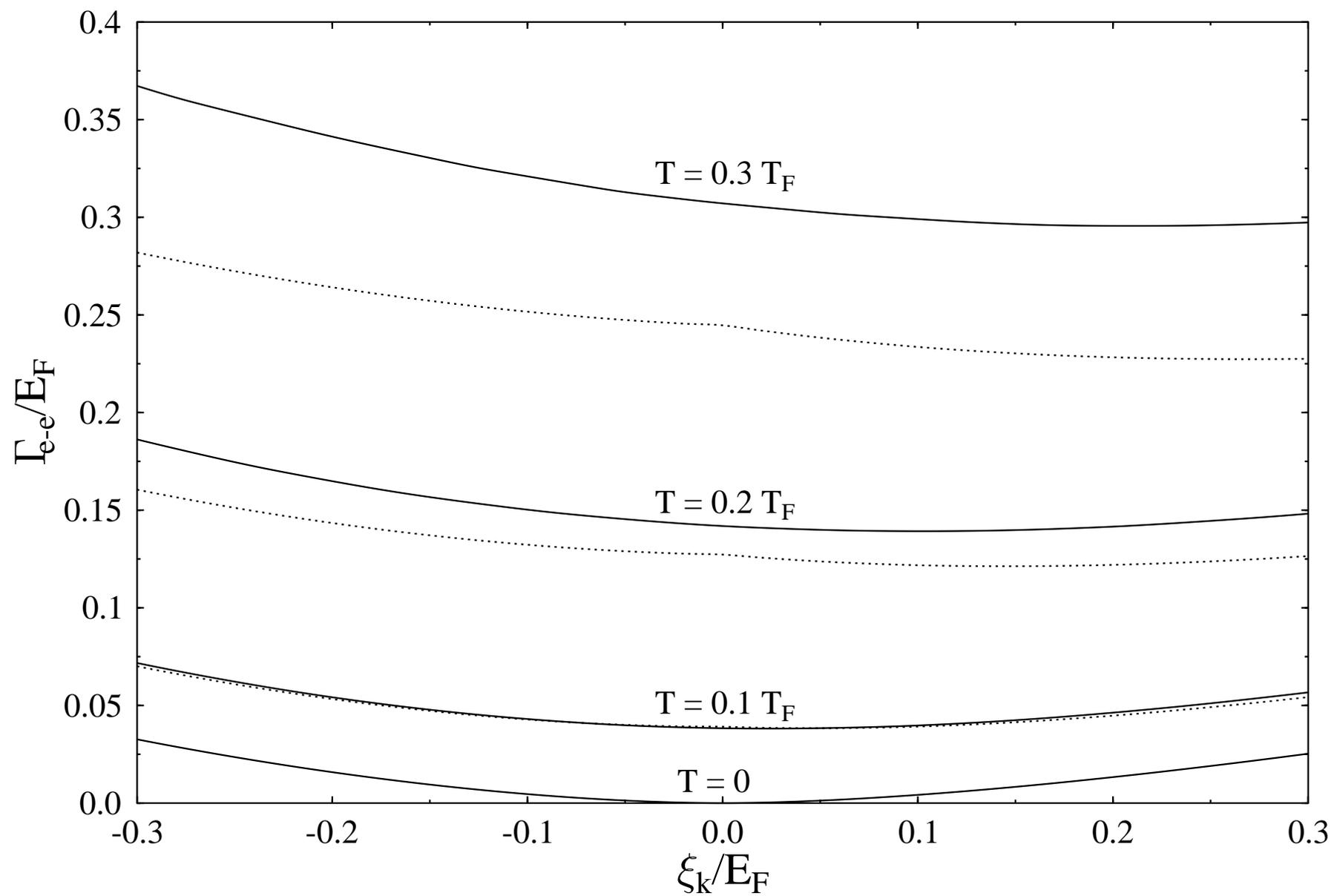
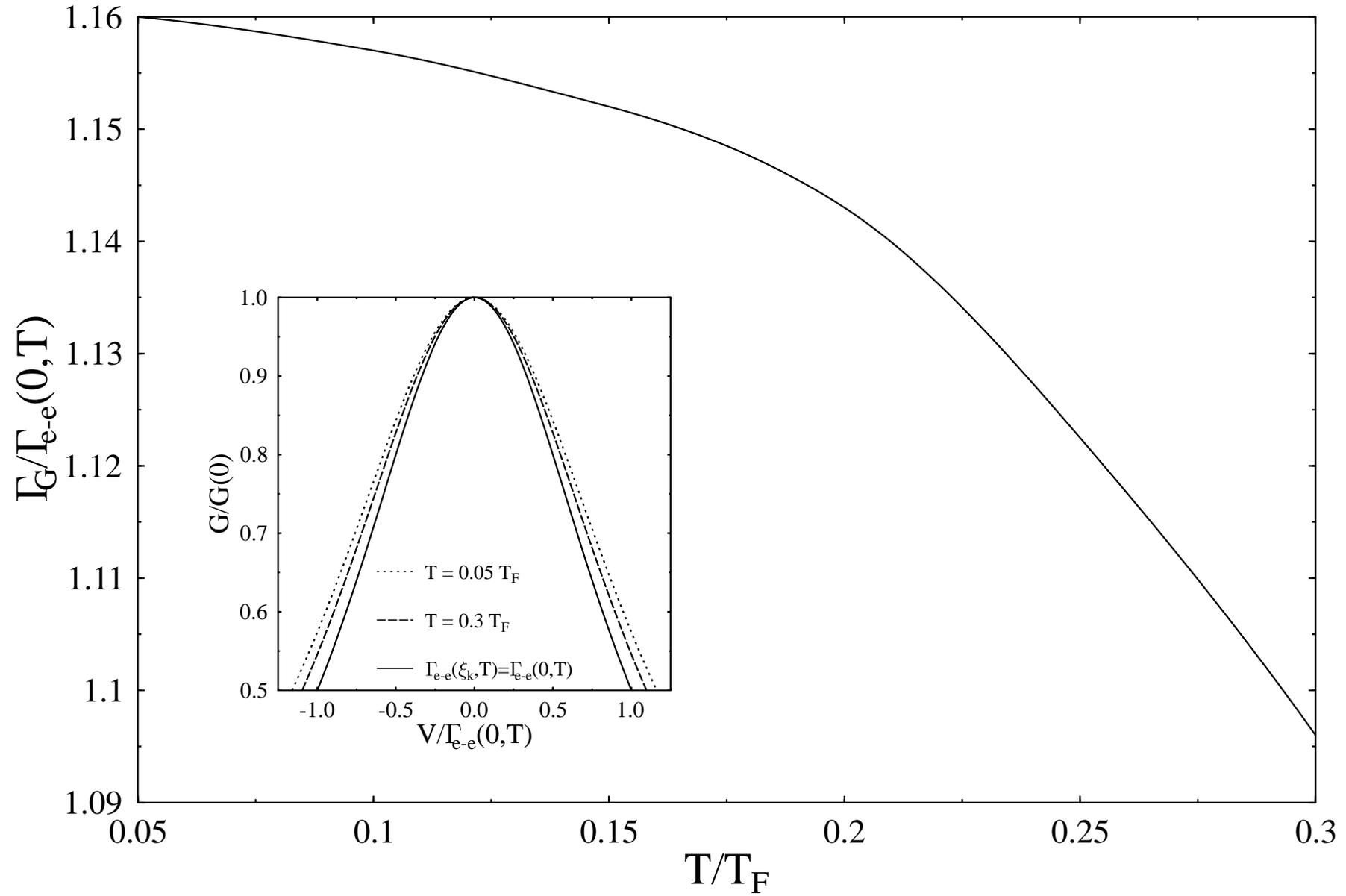
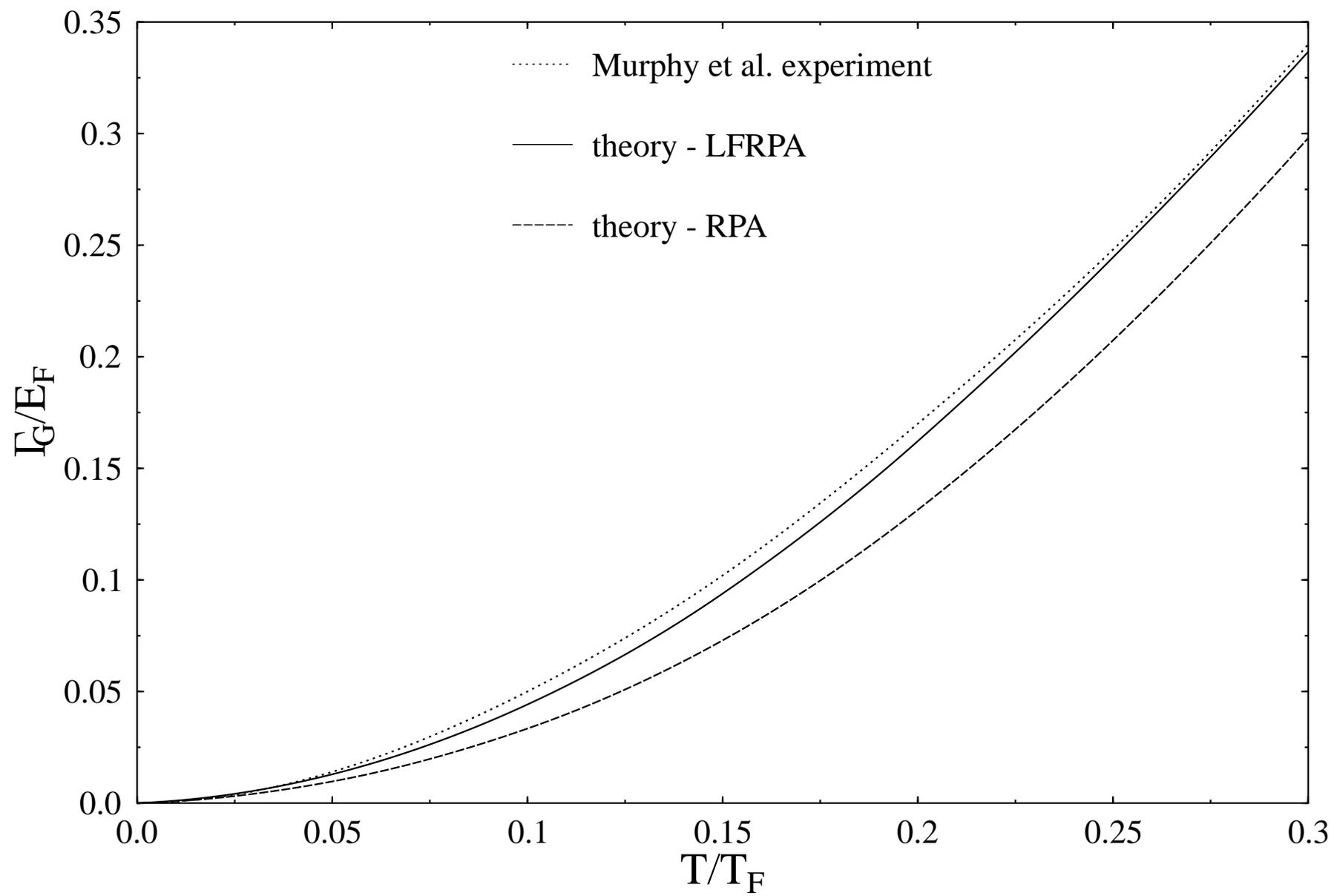
\end{document}